\documentclass{article} 
\usepackage{nips15submit_e,times}
\usepackage{hyperref}
\usepackage{url}
\usepackage{graphicx}
\usepackage{amsmath}
\usepackage{amssymb}
\usepackage{subfig}

\title{Accessing accurate documents by mining auxiliary document information }
\author{
Jinju Joby P. \\
Department of Computer Science and Engineering \\
Christ University Faculty of Engineering \\
Bangalore, India \\
\And
Jyothi Korra\\
Computer Science and Engineering\\
Christ University Faculty of Engineering\\
Bangalore, India\\
\texttt{jyothi.korra@christuniversity.in} \\
}

\nipsfinalcopy 
\begin{document}
\maketitle

\begin{abstract}
Earlier techniques of text mining included algorithms like k-means, Naïve Bayes, SVM which classify and cluster the text document for mining relevant information about the documents. The need for improving the mining techniques has us searching for techniques using the available algorithms. This paper proposes one technique which uses the auxiliary information that is present inside the text documents to improve the mining. This auxiliary information can be a description to the content. This information can be either useful or completely useless for mining. The user should assess the worth of the auxiliary information before considering this technique for text mining. In this paper, a combination of classical clustering algorithms is used to mine the datasets. The algorithm runs in two stages which carry out mining at different levels of abstraction. The clustered documents would then be classified based on the necessary groups. The proposed technique is aimed at improved results of document clustering.
\end{abstract}

\section{Introduction}

Text mining has reached greater levels as continuous research has been moving on to find newer and improved techniques to mine text data. The researchers have been using the basic mining algorithms and techniques such as K-means, agglomerative hierarchical clustering and such to find the best text mining results. The combinations of the techniques also have made quite an impressive mark in data mining like bisect-K-means \cite{ref3}. The commonly used clustering techniques like scatter/gather has been beneficial to understand how clustering happens in a real web world scenario \cite{ref8}. The available tools for text mining are helpful as most of them are open-source and have a wide range of options to work with. Some of the commercially available tools like SPSS are common to data miners \cite{ref2}. Apart from all these advances, researches are on an ever-growing thirst to find better ways to overcome the data mining difficulties as the data of the present world provides us with a range of parameters which cannot be well assessed using the existing tools and techniques. In such a search text mining on documents also require attention. The most important asset anywhere is data. To maintain the quality of data we go to extends of providing ways in which data can be managed, stored and used efficiently. As the amount of data increases exponentially as we speak, better ways to manage it is on high demand. Algorithms and techniques that do the same are to be found and deployed.
Text mining on documents has been done for years. Ways to mine the text and cluster the documents for better processing is our concern. The document mining algorithm that is proposed in this paper is a combination of few of the best text clustering algorithms and other data mining techniques. After the documents have been clustered into the most appropriate clusters, they are then classified into classes under which they belong most appropriately. The use of such document mining techniques can be applied in dataset management, to maintain data quality. The proposed techniques can be applied in order to cluster and classify the large number of documents that are otherwise disorganized. This is mainly required for easy access to the accurate document in minimum time. Thus, improving the mining techniques that can be used in the ever growing size of documents collected.

\section{Related Work}
The concept of document clustering has been a hot area of for research since the time data collection has taken birth. Researchers have been developing best and improved text mining algorithms and techniques since years. Even today as the size of data increases exponentially, text mining is a very important research area. This need has led to the proposal of an improved technique to mine the document dataset in order to cluster and classify the documents and thus to retrieve the most accurate document that is related to the query. The documents which are to be clustered are collected from a source which would be the World Wide Web or a repository which contains the different documents all in the same format \cite{ref10}\cite{ref11}. The issue of clustering these documents into a way which would trigger better retrieval should be developed. The raw form of these documents is just too scattered and would take a user hours or even days to search for the required document. Since processing time is very expensive in application it is the basic need to be clustered and classified for ease of use. The algorithms for text clustering and classification like Naïve Bayes, K-means, and SVM with kernerls \cite{dg1,dg2,dg3,} have been used to mine textual data \cite{ref1}. Naïve Bayes theorem was previously used to develop a system which detects the offensive content on the World Wide Web \cite{ref9}. The messages posted on the user walls were passed through the mining algorithm which runs based on the concept of Naïve Bayes. The resultant set of messages was classified based on the content and thus the messages with the offensive content would be blocked. The K-means algorithm is used strictly for clustering. The content based clustering requires an elaborate use of techniques along with K-means in order for the proposed methods to produce efficient results \cite{ref3}\cite{ref4}. The different techniques that are used for clustering and classification were summarized by various authors \cite{ref5}\cite{ref6}\cite{ref7}.
A lot of comparison is done between the algorithms for document clustering like agglomerative and K-means. K- means algorithm is used as it gives efficient results and the K- means algorithm gives more superior results. Scatter/Gather technique involves the hybrid working model of both K-means and agglomerative hierarchical clustering. The bisecting K- means also produces results of clustered documents that are as good as those developed by using the agglomerative hierarchical clustering\cite{ref3}. K-means is also applied after reducing the dimensions using PCA\cite{dg6}. 
The approach of using auxiliary information that is available along with the documents for the mining of documents is also a well-known approach. The web logs, provenance information that is attached along with the document as the side information. This technique would allow better text mining as the documents would be clustered into better clusters. The K-means algorithm is used to cluster the document into the initial clusters which are just vague clustering. Then the similarity measures are taken in order to cluster the documents better. Later the auxiliary information is made use of to improve the clustering and thus the resultant clustering of the algorithm would give us the refined clusters. This technique is extended to classify the documents and we would obtain clusters which will allow the fast retrieval of a query asking for a document from the large dataset.

\section{DOCUMENT CLUSTERING WITH AUXILARY INFORMATION}
The documents used in the clustering process should all contain auxiliary information. The second part of the proposed method as discussed in section IV would be used based on the auxiliary information. The first stage of clustering requires the documents to be clustered first based on K-means algorithms. The similarity between the documents is found using cosine similarity. The cluster centroids will be updated based on the similarity values. Now to improve the clustering process, we need to remove the noisy attributes which will be useless for the clustering. This can be done by calculating the Gini index of the attributes in the documents.

The set of useful attributes would be used for clustering thus refining the clustering process. Now the attributes and the document clusters have been selected. Now the proposed model would need to combine both these features in order to find the final clusters which will allow easier retrieval of documents. This can be done by finding the posterior probability of the documents with respect to the attributes and the clusters. The process would require the documents to have similar attributes so that the clustering and the calculation of Gini index can be easily dealt with. The designed algorithm that is explained in section IV would put light on how the scattered documents would be clustered and these clustered can be further classified into classes. These classes would be then labelled accordingly so that the system would be able to retrieve the accurate documents in the least execution time. Thus the documents that are clustered and classified based on the attributes or auxiliary information is found to return better results than the normal document clustering algorithms and techniques. Certain preprocessing techniques like removal of stop words and if the required attributes can be found then it should be arranged in the required format.
\begin{figure}
\centering
\includegraphics{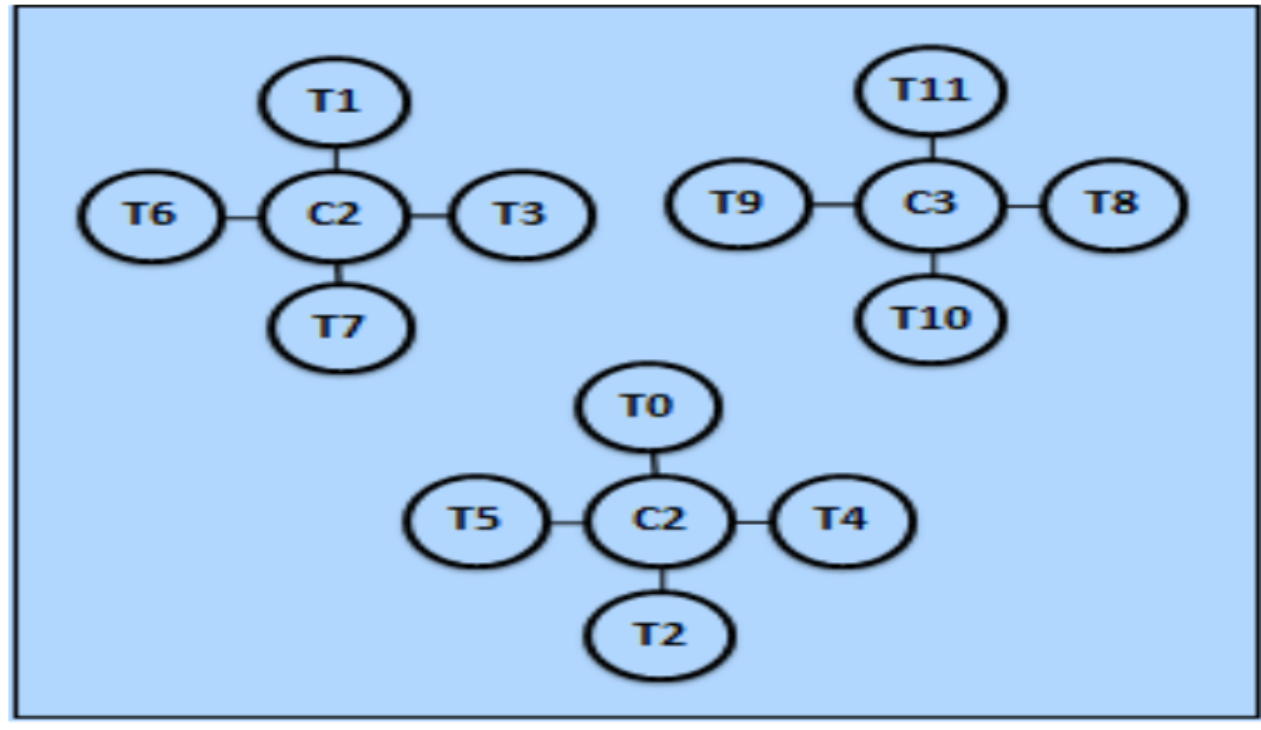}
\end{figure}
\section{PROPOSED METHOD}
In this section, we discuss the proposed method for the design on a clustering and classification technique for fast retrieval of documents from a large dataset. The entire process has been broken down for better grasping of the approach.
\subsection{PreprocessingDocuments}
The documents that are collected from the World Wide Web or repositories need not always have the same formatting. Also if the auxiliary information is known but is not incorporated in the document itself then it should be done. The auxiliary attributes should be distinguished from the rest of the content by appending a special character in front of it like a \$ or a \#. This way the algorithm can distinguish between the actual content of the document and the auxiliary information. This is important because different parts of the designed algorithm are used on the content and the auxiliary information separately. It is also well studied that information in web page is viewed in the order of position of words\cite{dg5}.
\subsection{Process Flow}
The process flow of the clustering process occurs as shown in Fig 2. The documents that are used for clustering purposes are always represented as vectors. Each document d would be considered as a vector d, the document can be expressed as $$df_t = tf_1 + tf_2 + ... +tf_n$$
, where dtf is the document and tfi is the frequency of the term i in the document.
The corpus would be all the documents in the dataset represented by T. The K-means algorithm is used to randomly select k documents as the cluster centroids and then the remaining documents are placed in any of the k clusters randomly. This way the first step of clustering is done. Now in order to find if the documents in the cluster are indeed correlated to each other, we need to check the similarity measure of the documents with the centroid of the cluster. The cosine similarity can be used for this checking. The similarity between document Ti and the cluster centroid Ci will determine if the document indeed belongs to that cluster or not. This way the cosine similarity of all the documents with all the centroids should be checked in order to make sure that the documents are present in the cluster with maximum relevance and is not misplaced. As we are still in the K-means algorithm, the next set of cluster centroids have to be calculated according to the new clusters formed. These clusters would be index on each document to mark them as clustered.
\begin{figure}
\centering
\includegraphics{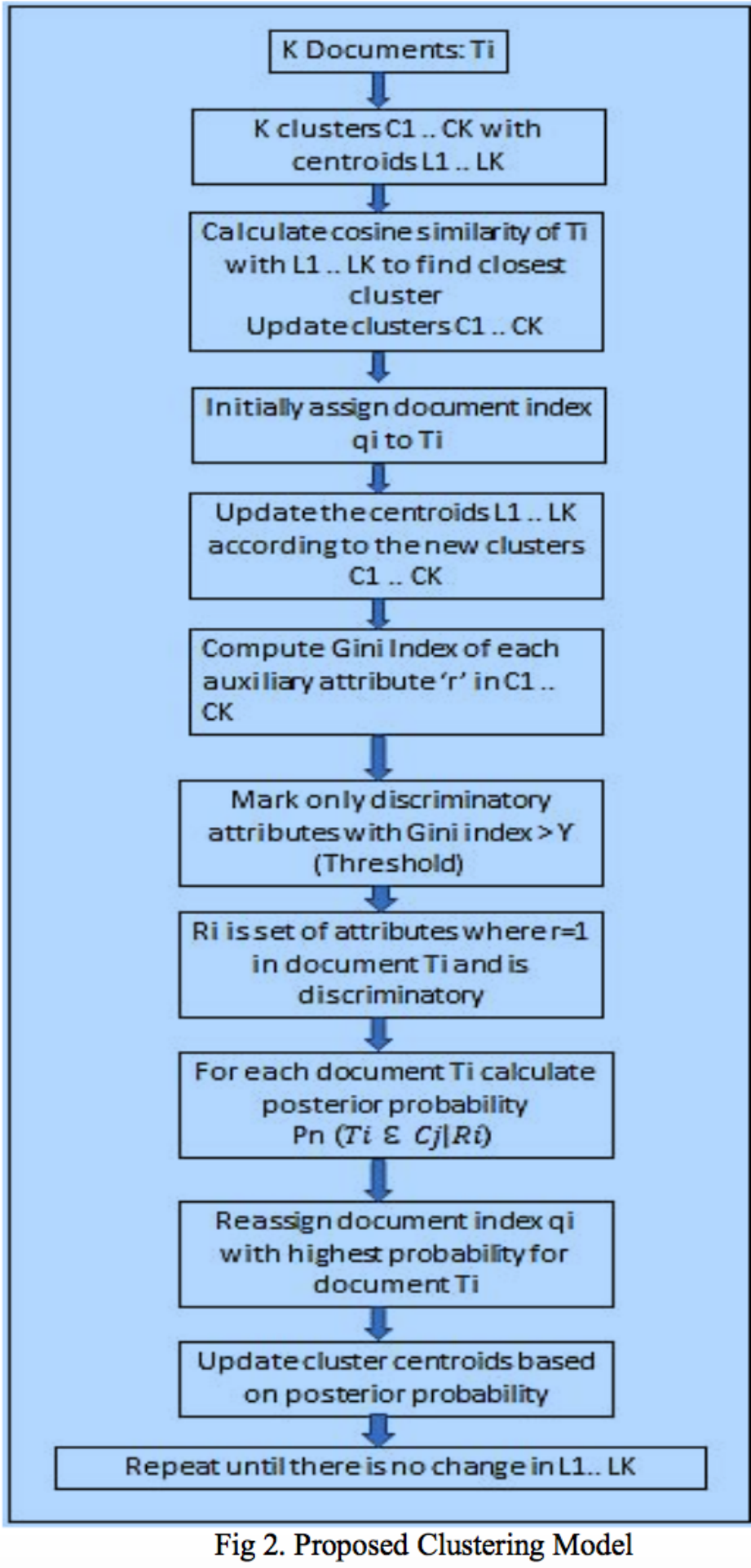}
\end{figure}
The part of clustering which involves only the documents is now done. Next we consider the auxiliary information or attribute that is present in the document. For example for the book ?In the year 2889?, the auxiliary information would be, \\
Title: In the Year 2889 \\
Author: Jules Verne and Michel Verne \\
Release Date: January 2, 2007 \\
Language: English \\
Character set encoding: ASCII. \\
The Gini index of each attribute r in the cluster Ci is found. The Gini index is calculated by
$$Pr_j = \frac{Fr_j}{\sum_{k=1}^n Fr_k}$$
, where Prj is the presence on attribute r in cluster j which is determined by the fraction of attribute r in cluster j, frj. If the attribute r is present in the cluster j then its value is 1.
The summation of all the Prj of the attribute r in all the clusters is the Gini index of the attribute. Higher the value of Gini index, more useful the attribute is. As we may consider useless attributes for the clustering process and this may reduce the effectiveness of the proposed algorithm, the useless attributes need to be trimmed away. Thus now we mark only those attributes that have Gini index above a threshold value as usable attributes. This will now contain only the set of attributes Ri that are absolutely useful for the clustering purpose. The correlation between the documents in the cluster and the most relevant attributes in the cluster is used to find the final cluster. This can be done by finding the posterior probability between the attributes and the documents in the cluster using, $$P(T_i \in C_i|R_i)$$
, which describes the posterior probability of attribute set Ri in the document Ti which is in the cluster Ci. The posterior probability of the document in each cluster is calculated and the highest value would be the cluster to which it belongs finally. This is the best cluster assignment that can be done based on the auxiliary attributes to the document.
\begin{figure}
\centering
\includegraphics{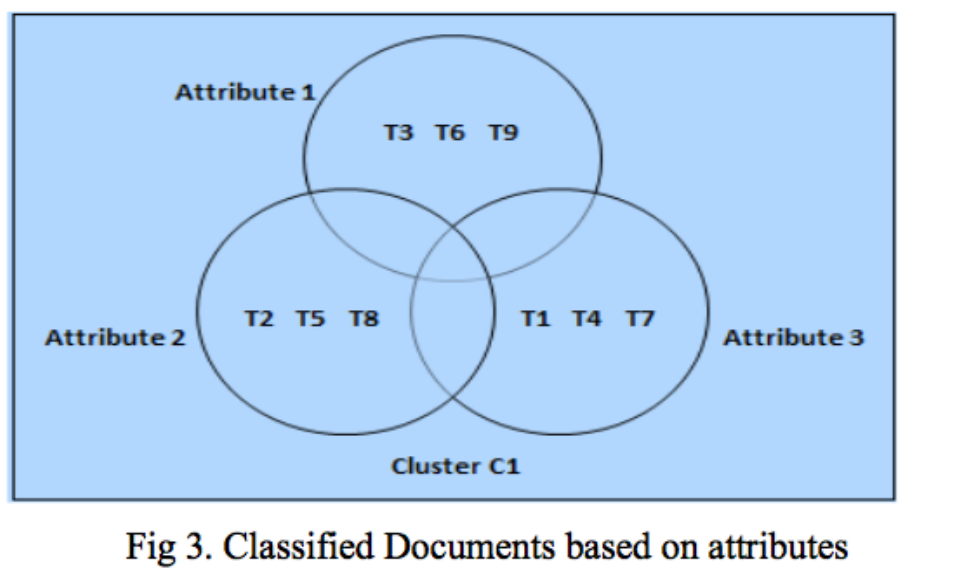}
\end{figure}

\subsection{Document Classification}
This section is an extension of the clustering process where the clusters that are obtained would be classified. In order to classify the clusters we require the knowledge of the type of documents that are present in each cluster. Classification accuracy is verified using crowdsourcing approach \cite{dg4}.
The documents that are clustered into a single cluster must contain maximum number of common attributes. This is because the clustering has been done based on only the most relevant attributes. Hence if the comparison is done between all the attributes in the cluster the type of majority documents could be obtained. This can also mean that the same cluster can be classified in many ways depending on the attributes present. One way of classification is by finding the most repeating attribute in the cluster and naming the cluster by the attribute. Another way of clustering is by calculating the presence of all the attributes in the cluster and the cluster will be named after the attribute whose presence would have the highest value. Presence of an attribute is calculated by the formula described in subsection B. The classification process flow is as shown in Fig 4. After classification the documents can belong to different classes because the different attributes can influence the classification as shown in Fig 3.
\begin{figure}
\centering
\includegraphics{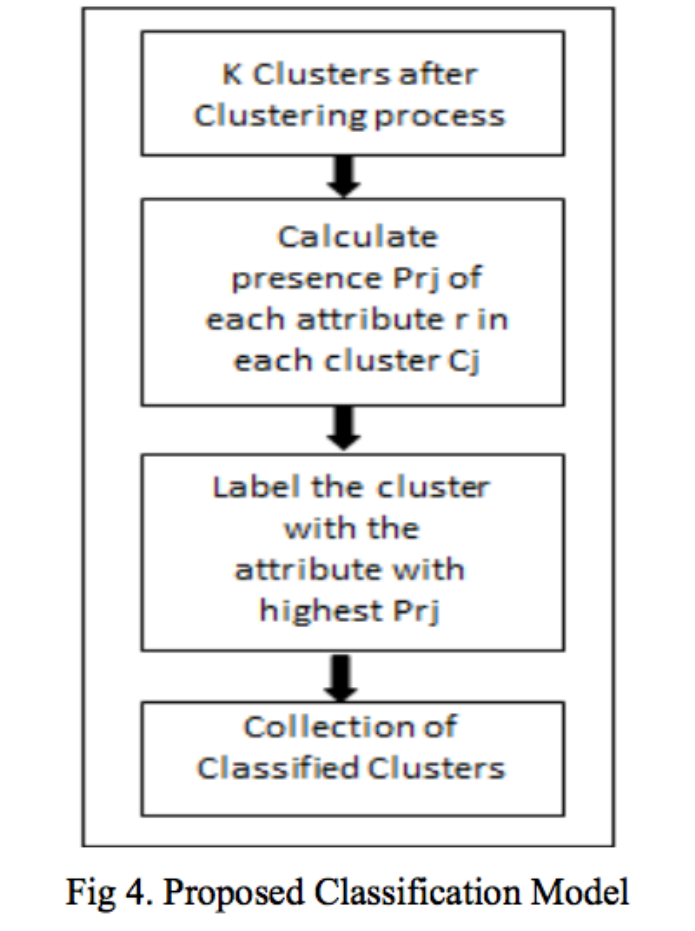}
\end{figure}

\section{EXPERIMENTAL RESULTS}
We have collected the datasets which contain a large set of documents. These documents are collected from a journal which contains the papers published in computer science and its applied areas. The algorithm is implemented into a system which takes the input as the dataset and the output would be in two stages. The correctly clustered set of documents also the classified clusters based on the highest attribute presence. Table I gives the details of the clustered documents and the cosine similarity values after the entire algorithm has processed the dataset. The above documents also have the attribute list as shown previously in section IV. The attributes would be useful to filter out the unnecessary steps of processing by removing the unnecessary and useless attributes. Thus the refined set of attributes only would be allowed for the further clustering of the documents. This is how the retrieval process becomes better when compared to the basic clustering algorithms and techniques. This clustering would contain the content and attribute based documents which would generate the most accurate documents for the query submitted by the user. This is the aim of the proposed algorithm.
\begin{figure}
\centering
\includegraphics{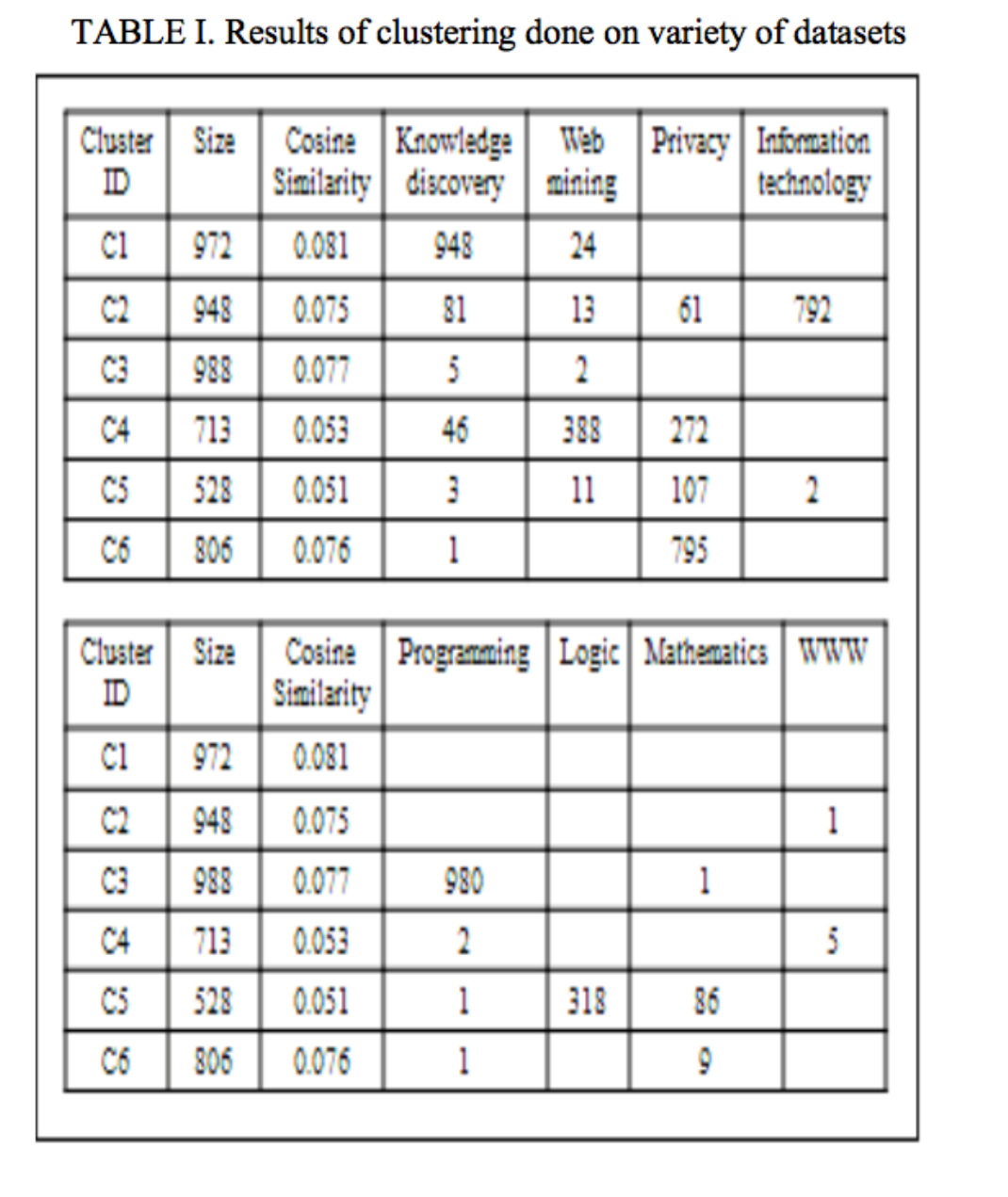}
\end{figure}

\section{CONCLUSION AND FUTURE WORK}
The proposed technique is aimed at the retrieval and accessing the documents out of a large dataset using the auxiliary attributes. The usual clustering process only uses the content in the documents to cluster. This paper proposes an approach to do the same, but by refining the clustering process using the attributes also.
The auxiliary attributes like author details, paper details and Web logs would be considered as auxiliary attributes. After clustering, the classification can be done by using the next step, where the most prominent attribute would be considered as the class of the cluster. This paper proposes this method in order to overcome the problem of maintaining and handling the large sets of data that we need to deal with.

{\small
\bibliographystyle{ieee}
\bibliography{egbib}
}

\end{document}